\documentclass[aps,pre,showpacs,amsmath,amssymb,twocolumn,10pt]{revtex4-1} 
\usepackage{amsthm}
\usepackage{dsfont}
\usepackage{psfrag}
\usepackage[dvips]{graphicx}

\newcommand\NN{{\mathds{N}}}

\newcommand\RR{{\mathds{R}}}

\newcommand\dd{{\mathrm{d}}}
\newcommand\ee{{\mathrm{e}}}
\newcommand\ii{{\mathrm{i}}}

\DeclareMathOperator{\Bernoulli}{{\mathrm{B}}}
\DeclareMathOperator{\arctanh}{{\mathrm{arctanh}}}
\DeclareMathOperator{\arccoth}{{\mathrm{arccoth}}}

\newtheorem*{hdc}{Criterion}

\begin{document} 

\title{Stationary point approach to the phase transition\\ of the classical $XY$ chain with power-law interactions} 

\author{Michael Kastner} 
\email{kastner@sun.ac.za} 
\affiliation{National Institute for Theoretical Physics (NITheP), Stellenbosch 7600, South Africa} 
\affiliation{Institute of Theoretical Physics,  University of Stellenbosch, Stellenbosch 7600, South Africa}

\date{\today}
 
\begin{abstract}
The stationary points of the potential energy function $V$ of the classical $XY$ chain with power law pair interactions (i.\,e., interactions decaying like $r^{-\alpha}$ with the distance) are analyzed. For a class of ``spinwave-type'' stationary points, the asymptotic behavior of the Hessian determinant of $V$ is computed analytically in the limit of large system size. The computation is based on the Toeplitz property of the Hessian and makes use of a Szeg\"o-type theorem. The results serve to illustrate a recently discovered relation between phase transitions and the properties of stationary points of classical many-body potentials. In agreement with this relation, the exact phase transition potential energy of the model can be read off from the behavior of the Hessian determinant for exponents $\alpha$ between zero and one. For $\alpha$ between one and two, the phase transition is not manifest in the behavior of the determinant, and it might be necessary to consider larger classes of stationary points.
\end{abstract}


\maketitle


\section{Introduction}
\label{sec:phasetransitions}

As has long been known, the stationary points of a classical Hamiltonian function or potential energy function can be employed to calculate or estimate certain physical quantities of interest. Well-known examples include transition state theory \cite{Eyring:35} or Kramers's reaction rate theory for the thermally activated escape from metastable states \cite{Kramers40}, where the barrier height (corresponding to the difference between potential energies at certain stationary points of the potential energy function) plays an essential role. More recently, the noise-free escape from quasi-stationary states (i.\,e., metastable states whose lifetimes diverge with the system size) has been related to the presence of stationary points of marginal stability \cite{Tamarit_etal05}. Apart from studies of dynamical properties, stationary points have also been extensively used for estimating thermodynamical properties by means of the superposition approach \cite{StrodelWales08}.

Dynamical properties like the aforementioned ones are, as one might expect, not unrelated to the statistical physical behavior of a system. Accordingly, as worked out beautifully in \cite{CaPeCo:00}, properties of stationary points of the potential energy function \footnote{The authors of Ref.\ \cite{CaPeCo:00} discuss topology changes of constant-energy manifolds in phase space. Via Morse theory, such topology changes can be related to the stationary points of a sufficiently smooth potential energy function.} reflect in dynamical and statistical physical quantities simultaneously. This observation sparked quite some research activity, reviewed in \cite{Kastner:08}, with the aim of relating equilibrium phase transitions to stationary points and their indices. Most importantly, it was shown in \cite{FraPe:04,*FraPeSpi:07} that, under a number of technical conditions, the presence of stationary points of the potential energy function is necessary for a phase transition to take place. Subsequently, it was noticed in \cite{KaSchne:08,*KaSchneSchrei:08} that the Hessian determinant of the potential energy function, evaluated at the stationary points, adds a crucial piece of information for discriminating whether or not a phase transition occurs. Omitting some of the technicalities, the essence of the criterion on the Hessian determinant can be captured as follows \cite{NardiniCasetti09}.
\begin{hdc}
Let
\begin{equation}\label{eq:StandardHamiltonian}
H(p,q)=\frac{C}{2}\sum_{k=1}^N p_k^2 + V(q_1,\dots,q_N),
\end{equation}
with some constant $C\geqslant0$, be the total energy function of a system with $N$ degrees of freedom, where $p=(p_1,\dots,p_N)$ and $q=(q_1,\dots,q_N)$ denote the vectors of momenta and positions. The potential energy $V$ will, in general, have stationary points $q_\text{s}^N$ defined as solutions of the set of equations
\begin{equation}
0=\left.\frac{\partial V(q)}{\partial q_k}\right|_{q=q_\text{s}^N},\qquad k=1,\dots,N.
\end{equation}
The stationary points are assumed to be isolated, and their number is assumed to grow at most exponentially with $N$. In the thermodynamic limit $N\to\infty$, the stationary points can induce a phase transition at some critical energy per degree of freedom $e_\text{c}$ only if the following two conditions are met:
\begin{enumerate}
\item There exists a sequence $\bigl\{q_\text{s}^N\bigr\}_{N=N_0}^\infty$ of stationary points of $V$ such that
\begin{equation}\label{eq:vc_limit}
v_\text{c}:=\lim_{N\to\infty}\frac{V\bigl(q_\text{s}^N\bigr)}{N}
\end{equation}
converges and $v_\text{c}=\langle v\rangle(e_\text{c})$ is the ensemble expectation value of $v=V/N$ at the energy $e_\text{c}$.
\item The asymptotic behavior of the Hessian matrix $\mathcal{H}$ of $V$, evaluated at the critical points $q_\text{s}^N$ contained in that sequence, is such that
\begin{equation}\label{eq:criterion}
\lim_{N\to\infty}\bigl|\det \mathcal{H}\bigl(q_\text{s}^N\bigr)\bigr|^{1/N}=0.
\end{equation}
\end{enumerate}
\end{hdc}
In short, the criterion requires the existence of a sequence of stationary points whose potential energy converges to $v_\text{c}$ and whose Hessian determinant vanishes in the sense of \eqref{eq:criterion} in the thermodynamic limit. Setting the constant $C$ in \eqref{eq:StandardHamiltonian} to zero, the kinetic energy term is absent, as is the case for many classical spin models. The criterion therefore remains valid in this case, with the only differences that the total energy $H$ equals the potential energy $V$, and the critical energy $e_\text{c}$ is identical to the critical potential energy $v_\text{c}$ \cite{CaKaNe09}. This is the case we will be concerned with in the present article.

Note that the above criterion is not sufficient for a phase transition to occur: finding a sequence of stationary points with the behavior specified above does not guarantee a transition to take place at the corresponding critical energy. However, as model calculations suggest, the criterion usually appears to single out the correct transition energies \cite{KaSchne:08,*KaSchneSchrei:08}. Importantly for the application of the criterion, it is not necessary to know all stationary points of $V$, but a suitably chosen subset may be sufficient. This matter of fact was pointed out by Nardini and Casetti, and suitably constructed sequences of stationary points were used in \cite{NardiniCasetti09} to single out the phase transition of a model of gravitating masses and analytically determine its critical energy.

Comparing the Hessian determinant criterion to other analytic tools in the statistical physics of phase transitions, its remarkable property is that it is {\em local}\/ in configuration space. In contrast to, say, the calculation of a partition function, no averaging over a large, high-dimensional manifold is necessary. Instead, only the local properties of a sequence of stationary points need to be analyzed. Of course, finding an appropriate sequence of stationary points can be equally hard or impossible, but in certain instances such a local approach may prove beneficial.

In this article, we study the stationary points of the potential energy function of a chain of classical $XY$ spins (or rotators), coupled by a pair interaction which decays like $r^{-\alpha}$ with the distance $r$ of the spins on the lattice. The model is introduced in detail in Sec.\ \ref{sec:model}. Although one-dimensional, it shows a phase transition from a ferromagnetic to a paramagnetic phase for exponents $\alpha$ between zero and two, and the aim of the present work is to explore the relation between stationary points and phase transitions for these values of $\alpha$.

There are a number of interesting aspects of this study that deserve mention: First, inspired by Ref.\ \cite{NardiniCasetti09}, in Sec.\ \ref{sec:stationarypoints} a method is devised of how to construct special classes of stationary points for lattice spin systems. The potential energy at such stationary points is evaluated in Sec.\ \ref{sec:energy}. The Hessian at such a stationary point, as required by \eqref{eq:criterion}, is found to be a Toeplitz matrix. As carried out in Sec.\ \ref{sec:Hessian}, this property allows us to employ a Szeg\"o-type theorem for the calculation of the asymptotic behavior of the Hessian determinant in the limit of large system sizes $N$. The results of Sec.\ \ref{sec:Hessian} depend on the exponent $\alpha$ not only quantitatively, but also qualitatively: For $0\leqslant\alpha\leqslant1$, the asymptotic behavior of the Hessian determinant indeed signals the phase transition at the exact value of the transition energy, as purported by the criterion of Sec.\ \ref{sec:phasetransitions}. For $1<\alpha\leqslant2$, no signature of the phase transition is detected from the Hessian determinant of the special class of stationary points considered, and one might conclude that other (or larger) classes of stationary points have to be taken into account. The findings are summarized and discussed in Sec.\ \ref{sec:discussion}.



\section{Classical $XY$ chain with power-law interactions}
\label{sec:model}
Consider a set of $N$ lattice sites labeled by an integer number $j\in\{1,\dots,N\}$ where, to ease the notation, we assume $N$ to be odd. To each site, a planar vector of unit length is assigned, parametrized by the angular variable $\theta_j\in(-\pi,\pi]$. The classical $XY$ chain  with power law interactions is characterized by the potential energy function
\begin{equation}\label{eq:Hamiltonian}
V(\theta)=\mathcal{N}\sum_{i=1}^N\sum_{j=1}^{(N-1)/2} \frac{1-\cos(\theta_i-\theta_{i+j})}{j^\alpha}
\end{equation}
with $\theta=(\theta_1,\dots,\theta_N)$ and some nonnegative exponent $\alpha$. 
Although suppressed in the notation, indices $i$ of the $\theta_i$ variables are always to be considered modulo $N$, such as to account for periodic boundary conditions and to guarantee indices in the range from 1 to $N$.
The potential energy function \eqref{eq:Hamiltonian} describes $N$ classical spin variables on a ring (chain with periodic boundary conditions), where each spin interacts with every other. The interaction strength between two spins decays proportionally to $1/j^\alpha$, where $j$ is the minimal distance of the two spins on the ring. The potential energy \eqref{eq:Hamiltonian} is endowed with a normalization factor defined as
\begin{equation}\label{eq:normalization}
\mathcal{N}=\Biggl(2\sum_{j=1}^{(N-1)/2}\frac{1}{j^\alpha}\Biggr)^{-1}.
\end{equation}
The asymptotic behavior of $\mathcal{N}$ in the limit $N\to\infty$ can be computed, yielding
\begin{equation}\label{eq:Nasymptotic}
2\mathcal{N}\sim
\begin{cases}
(1-\alpha)2^{1-\alpha}N^{\alpha-1} &\text{for $0\leqslant\alpha<1$},\\
1/\ln N &\text{for $\alpha=1$},\\
1/\zeta(\alpha) &\text{for $\alpha>1$},
\end{cases}
\end{equation}
where $\zeta$ denotes the Riemann zeta function. This normalization factor, introduced in Ref.\ \cite{AnteneodoTsallis98}, is chosen such as to guarantee extensivity of the potential energy, i.e., a finite limit of the potential energy per particle in the limit $N\to\infty$.

The thermodynamic behavior of this model depends on the exponent $\alpha$ in the following way: For $0\leqslant\alpha\leqslant1$, the thermodynamic behavior is identical to that of the mean-field (or Curie-Weiss) case $\alpha=0$, showing a ferromagnetic continuous phase transition characterized by mean-field critical exponents \cite{TamaritAnteneodo00,*CampaGiansantiMoroni00}. For $1<\alpha\leqslant2$, the model also shows a phase transition, but thermodynamic functions differ from the mean-field case (see \cite{Kosterlitz76} and the comment in Sec.\ 5 of Ref.\ \cite{Froehlich_etal78}). For $\alpha>2$, no phase transition occurs. The three regimes for the exponent $\alpha$, the corresponding thermodynamic behavior, and also the methods of proof are analogous to Dyson's analysis of the Ising chain with spin-spin interaction strengths decaying as $1/j^\alpha$ \cite{Dyson69a,*Dyson69b}.

\section{Stationary points}
\label{sec:stationarypoints}

Stationary points of the potential energy function \eqref{eq:Hamiltonian} are defined as the real solutions of the set of equations
\begin{equation}\label{eq:statpoints}
\begin{split}
0&=\frac{\partial V(\theta)}{\partial \theta_k}\\
&= \mathcal{N}\sum_{j=1}^{(N-1)/2}\frac{\sin(\theta_k-\theta_{k+j})+\sin(\theta_k-\theta_{k-j})}{j^\alpha}
\end{split}
\end{equation}
for $k=1,\dots,N$. Since the potential energy function \eqref{eq:Hamiltonian} is invariant under a global rotation $\theta_i\to\theta_i+\phi$ with $\phi\in\RR$, the solutions of \eqref{eq:statpoints} come in one-parameter families: Given a stationary point $(\theta_1,\dots,\theta_N)$, every point $(\theta_1+\phi,\dots,\theta_N+\phi)$ is also a solution of \eqref{eq:statpoints}. The criterion of Sec.\ \ref{sec:phasetransitions}, however, requires all stationary points to be isolated, and we therefore have to get rid of the trivial rotational degeneracy. We explicitely destroy the rotational symmetry by fixing $\theta_N=0$ and eliminating the equation with $k=N$ in \eqref{eq:statpoints}. The thermodynamics of this reduced model is identical to that of the full one, since the contribution of one degree of freedom to the partition function is negligible in the thermodynamic limit.

Determining all solutions of the remaining set of nonlinear equations is presumably a hard task (too hard for the author at least). There are, however, two particularly simple classes of solutions, similar in spirit to those constructed in \cite{NardiniCasetti09} for a one-dimensional model of gravitating masses: First, any combination of $\theta_i\in\{0,\pi\}$ for $i=1,\dots,N-1$ will make the sine functions in \eqref{eq:statpoints} vanish and therefore satisfies the set of equations. A second class of solutions is given by
\begin{equation}\label{eq:isoangular}
\theta_m^{(x)}=mx\qquad\text{with $x=2\pi n/N$},
\end{equation}
where $m,n\in\{1,\dots,N\}$ and hence $0<x<2\pi$. These solutions have constant radian $x$ between neighboring spins, implying $\sin(\theta_k-\theta_{k+j})=\sin(\theta_{k-j}-\theta_k)$ and therefore each of the summands in \eqref{eq:statpoints} vanishes separately.

\begin{figure}\center
\includegraphics[width=0.85\linewidth]{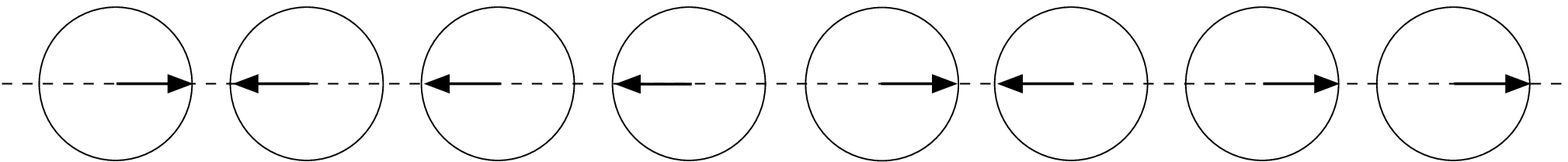}\\[5mm]
\includegraphics[width=0.85\linewidth]{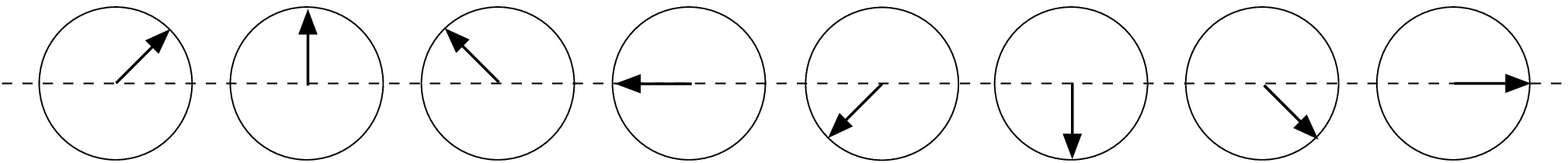}
\caption{\label{fig:classes}
Sketch of stationary points of the potential $V$ for $N=8$, where $\theta_i$ is the angle between the arrow and the dashed axis. Top: Stationary points where all $\theta_i\in\{0,\pi\}$. Bottom: Spinwave stationary point \eqref{eq:isoangular}, where all differences $\theta_i-\theta_{i-1}$ between neighboring angles are equal, with differences chosen such that $\theta_0=\theta_N$, in compliance with the periodic boundary conditions.
}
\end{figure}

These two classes of solutions are illustrated in Fig.\ \ref{fig:classes} and, as is easily checked numerically, are not exhaustive. This is probably expected, in particular when comparing with the results for the nearest-neighbor $XY$ chain for which all stationary points can be computed analytically \cite{MehtaKastner}. The two classes of solutions introduced above are also solutions in the case of nearest-neighbor interactions, but many more solutions exist. Since nearest-neighbor interactions can be considered as the limit $\alpha\to\infty$ of the power law decay discussed in the present article, it is maybe not too surprising to find that (at least many of) these solutions persist to finite $\alpha$.

We will in the following restrict the analysis to ``spinwave'' stationary points \eqref{eq:isoangular}, mainly for the reason that the Hessian matrix at these points, as discussed in detail in Sec.\ \ref{sec:Hessian}, is a Toeplitz matrix. This structure is particularly helpful when calculating the large-$N$ asymptotics of the Hessian determinant. Moreover, from the results on the nearest-neighbor $XY$ chain in \cite{MehtaKastner}, one may be led to conjecture that the spinwave stationary points are of particular importance for our purposes: At least for nearest-neighbor interactions, the spinwave stationary points have the smallest absolute value of the Hessian determinant amongst the stationary points of a given potential energy, and therefore determine whether the criterion of Sec.\ \ref{sec:phasetransitions} is satisfied or not.

\section{Potential energy at stationary points}
\label{sec:energy}

The criterion of Sec.\ \ref{sec:phasetransitions} involves the potential energy evaluated at the stationary points. Inserting the spinwave stationary points \eqref{eq:isoangular} into the potential energy function \eqref{eq:Hamiltonian}, we obtain
\begin{eqnarray}
v(x)&:=&\frac{V(\theta^{(x)})}{N}=\frac{\mathcal{N}}{N}\sum_{i=1}^N\sum_{j=1}^{(N-1)/2} \frac{1-\cos(jx)}{j^\alpha}\nonumber\\
&=&\frac{1}{2}-\mathcal{N}\sum_{j=1}^{(N-1)/2} \frac{\cos(jx)}{j^\alpha}.\label{eq:energy}
\end{eqnarray}
Since $|\cos x|\leqslant1$ for all $x$, we have
\begin{equation}
\Biggl|\mathcal{N}\sum_{j=1}^{(N-1)/2} \frac{\cos(jx)}{j^\alpha}\Biggr|\leqslant \mathcal{N}\sum_{j=1}^{(N-1)/2} \frac{1}{j^\alpha}=\frac{1}{2},
\end{equation}
confirming that the normalization factor $\mathcal{N}$ in \eqref{eq:normalization} had been chosen appropriately in order to render the potential energy per spin finite in the thermodynamic limit.

\begin{figure}\center
\psfrag{e}{$v$}
\psfrag{xxx}{$x/\pi$}
\includegraphics[width=0.95\linewidth]{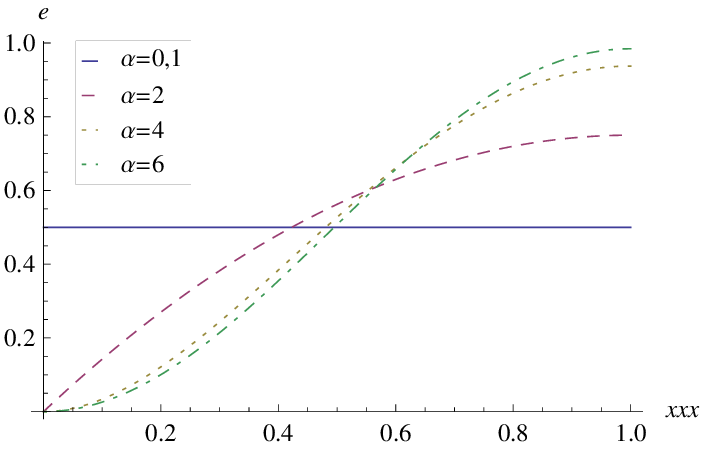}
\includegraphics[width=0.95\linewidth]{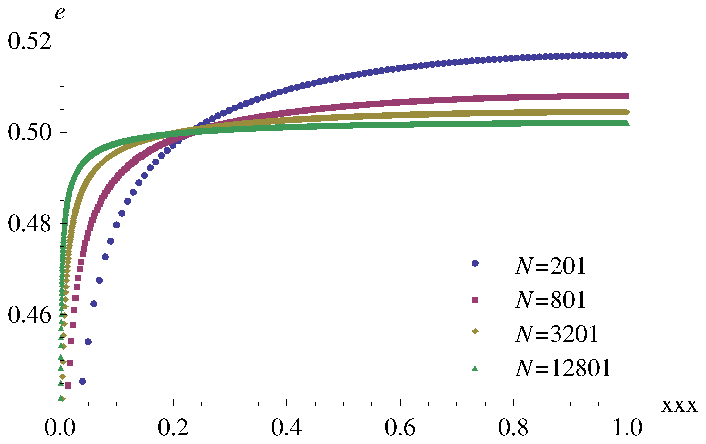}
\caption{\label{fig:e_of_x}
(Color online) Upper plot: The graph of the potential energy per spin $e(x)$ of a spinwave solution $\theta^{(x)}$ in the thermodynamic limit as given in \eqref{eq:e_of_x}. For a given positive, even $\alpha=2,4,6,\dots$, the energy values cover densely the entire range of energies per spin accessible to the system in the thermodynamic limit. Lower plot: Potential energy per spin $e(x)$ for $\alpha=1/2$ and spinwave stationary points \eqref{eq:isoangular}, plotted for various system sizes $N$. With increasing $N$, the curve approaches a horizontal line of energy $1/2$.
}
\end{figure}

In the limit $N\to\infty$ and for certain values of the exponent $\alpha$, the summation in the second line of \eqref{eq:energy} can be performed explicitely: For $\alpha=1$ we use formula 1.441.2 of \cite{GradshteynRyzhik}, and for $\alpha\in2\NN_0$ formula 1.443.1 of the same reference, to obtain
\begin{equation}\label{eq:e_of_x}
v(x)=\frac{1}{2}+\frac{\mathcal{N}}{2}
\begin{cases}
\ln[2(1-\cos x)] & \text{for $\alpha=1$},\\
(-4\pi^2)^{\alpha/2}\Bernoulli_\alpha[x/(2\pi)]/\alpha! & \text{for $\alpha\in2\NN_0$},
\end{cases}
\end{equation}
for the potential energy of a spinwave solution $\theta^{(x)}$ in the thermodynamic limit. $B_\alpha$ denotes the Bernoulli polynomial of order $\alpha$ as defined, for example, in Sec.\ 9.62 of \cite{GradshteynRyzhik}. The graph of $v(x)$ is shown in Fig.\ \ref{fig:e_of_x} (upper plot) for exponents $\alpha=0,1,2,4$, and 6. For noninteger values of $\alpha$, the infinite sum in \eqref{eq:energy} cannot be performed, but $v(x)$ can be evaluated numerically for reasonably large system sizes $N$. The resulting curves (not shown in Fig.\ \ref{fig:e_of_x}) are found to interpolate smoothly between the curves for integer $\alpha$. For a given positive, even $\alpha=2,4,6,\dots$, 
the potential energy values cover densely the entire range of potential energies per spin accessible to the system in the thermodynamic limit. This is a desirable property when applying the criterion of Sec.\ \ref{sec:phasetransitions}, as it allows us to use spinwave stationary points for the construction of sequences of stationary points whose potential energies converge to any desired value accessible to the system.

For exponents in the range $0\leqslant\alpha\leqslant1$, the situation is more intricate. For large, but finite, system sizes $N$, the potential energies corresponding to the spinwave stationary points 
$\theta^{(x)}$ become denser and denser on the interval $(0,1/2)$ with increasing $N$. In the thermodynamic limit, however, the potential energy converges to $1/2$ for any given value of $x\in(0,2\pi)$, and to zero for $x=0$, resulting in the straight line plotted in Fig.\ \ref{fig:e_of_x} (upper plot). The approach to this behavior with increasing system size $N$ is illustrated in Fig.\ \ref{fig:e_of_x} (lower plot) for the exponent $\alpha=1/2$.

Equation \eqref{eq:e_of_x} and the numerical results in Fig.\ \ref{fig:e_of_x} demonstrate that the potential energy per spin of a sequence $\{\theta^{(x)}\}$ of spinwave stationary points for increasing $N$, but with a fixed value of $x$, indeed converges to a limiting value, as required in equation \eqref{eq:vc_limit} of the criterion in Sec.\ \ref{sec:phasetransitions}. To construct a sequence of stationary points with a given value of $x$, it will in general be necessary to restrict the sequence to some infinite subset $\{N_1,N_2,\dotsc\}$ of system sizes such that $2\pi n/N_i=x$ for some $n\in\{1,\dotsc,N_i\}$.


\section{Hessian determinant at stationary points}
\label{sec:Hessian}

In order to apply the criterion stated in Sec.\ \ref{sec:phasetransitions}, we also need to evaluate the determinant of the Hessian matrix at a stationary point. Again, in order to destroy the trivial global rotational symmetry of the potential energy function $V$, one of the spin variables, say, $\theta_N$, is fixed at zero. The resulting potential energy is a function of $N-1$ variables $\theta_1,\dots,\theta_{N-1}$, and its Hessian $\mathcal{H}_N$ is an $(N-1)\times(N-1)$ symmetric matrix with entries
\begin{widetext}
\begin{equation}
[\mathcal{H}_N]_{kl}(\theta)=\frac{\partial^2 V(\theta)}{\partial \theta_k \partial \theta_l}=
\begin{cases}
\displaystyle
\mathcal{N}\sum_{j=1}^{(N-1)/2}\frac{\cos(\theta_k-\theta_{k+j})+\cos(\theta_k-\theta_{k-j})}{j^\alpha} & \text{for $k=l$},\\
\displaystyle
-\mathcal{N}\frac{\cos(\theta_k-\theta_l)}{\Delta(l-k)^\alpha} & \text{for $k\neq l$},
\end{cases}
\end{equation}
\end{widetext}
for $k,l=1,\dots,N-1$, where
\begin{equation}
\Delta(l-k)=
\begin{cases}
|l-k| &\text{for $|l-k|\leqslant (N-1)/2$},\\
N-|l-k| &\text{else},
\end{cases}
\end{equation}
is the minimal distance between $k$ and $l$ on the ring. Evaluating the Hessian at a spinwave stationary point $\theta^{(x)}$ as defined in \eqref{eq:isoangular}, one obtains
\begin{equation}\label{eq:Hthetan}
[\mathcal{H}_N]_{kl}(\theta^{(x)})=
\begin{cases}
\displaystyle
1-2v(x) & \text{for $k=l$},\\
\displaystyle
h_{l-k}^{(x)} & \text{for $k\neq l$},
\end{cases}
\end{equation}
with
\begin{equation}\label{eq:h_j}
h_j^{(x)}=-\mathcal{N}\frac{\cos(jx)}{\Delta(j)^\alpha}
\end{equation}
and with the potential energy per spin $v(x)$ as given in \eqref{eq:energy}. Without fixing $\theta_N$ to zero, this matrix would be circulant and the eigenvalues were readily obtained by Fourier transforming a row vector of the matrix. Fixing $\theta_N$ corresponds to eliminating the $N$th row and column of the matrix, and although the resulting matrix is not circulant anymore, it retains the Toeplitz property: As is evident from \eqref{eq:Hthetan}, the elements $[\mathcal{H}_N]_{kl}$ depend only on the difference $l-k$ of the indices.

\subsection{{S}zeg\"o's theorem}
\label{sec:Szego}

For our purposes, the Toeplitz property comes in handy, as a number of theorems on the large-$N$ asymptotics of determinants are known for sequences of $N\times N$ Toeplitz matrices \cite{BoettcherSilbermann}. The kind of sequence $\{T_N\}_{N=1}^\infty$ of matrices $T_N$ that is typically considered in the mathematics literature is where the matrix elements
\begin{equation}
[T_N(f)]_{kl}\equiv t_{l-k}(f)
\end{equation}
are given as Fourier coefficients of a complex-valued function $f$ defined on the circle,
\begin{equation}\label{eq:T_of_f}
t_j(f)=\frac{1}{2\pi}\int_{-\pi}^{\pi}f(\phi)\ee^{-\ii j\phi}\dd\phi.
\end{equation}
For particularly well-behaved $f$, Szeg\"o's theorem states that the large-$N$ asymptotic behavior of the determinants of such a sequence is given by
\begin{equation}\label{eq:Szego}
\lim_{N\to\infty}\bigl|\det\bigl(T_N(f)\bigr)\bigr|^{1/N}=\exp\left(\frac{1}{2\pi}\int_0^{2\pi}\ln f(\phi)\dd\phi\right),
\end{equation}
and many generalizations of this result to larger classes of symbols $f$ can be found in the literature \cite{BoettcherSilbermann}. 
Inverting the Fourier transformation \eqref{eq:T_of_f}, we can write
\begin{equation}
f(\phi)=\sum_{j=-\infty}^\infty t_j\ee^{\ii j\phi}.
\end{equation}
For the $XY$ chain, using equations \eqref{eq:Hthetan} and \eqref{eq:h_j} and a standard trigonometric identity, we obtain
\begin{equation}\label{eq:fofx}
\begin{split}
f^{(x)}(\phi)=&1-2v(x)\\
&-\mathcal{N}\sum_{j=1}^\infty\frac{\cos[j(x+\phi)]+\cos[j(x-\phi)]}{j^\alpha}.
\end{split}
\end{equation}
for the symbol of the Hessian, evaluated at a spinwave stationary point $\theta^{(x)}$ as defined in \eqref{eq:isoangular}. 
Then, as in the calculation of the potential energy in Sec.\ \ref{sec:energy}, the formul\ae\ 1.441.2 and 1.443.1 of \cite{GradshteynRyzhik} can be used to perform the summation in \eqref{eq:fofx} for the values $\alpha=1$ or $\alpha\in2\NN$ of the exponent.

\subsubsection{Exponent $\alpha=1$}
\begin{figure}\center
\psfrag{DDDDDDD}{$\left|\det\mathcal{H}_N\right|^{1/N}$}
\psfrag{e}{$v$}
\includegraphics[width=0.95\linewidth]{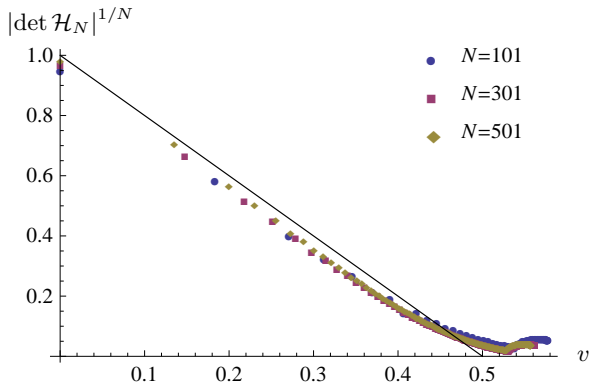}
\caption{\label{fig:D1_of_e}
(Color online) The $N$th root of the Hessian determinant $\det\mathcal{H}_N$ for $\alpha=1$, evaluated at spin wave stationary points $\theta^{(x)}$, plotted versus the corresponding potential energy per spin $v(x)$. With increasing system size $N$, the numerically computed determinant slowly (logarithmically) approaches the analytic large-$N$ asymptotic result $\mathcal{D}_1$ (black line).
}
\end{figure}

In the case of $\alpha=1$ we can use the identity
\begin{equation}
\sum_{j=1}^\infty \frac{\cos(jx)}{j}=-\frac{1}{2}\ln[2(1-\cos x)]
\end{equation}
(formula 1.441.2 of \cite{GradshteynRyzhik}) to write \eqref{eq:fofx} in the form
\begin{equation}\label{eq:f1}
f^{(x)}(\phi)=\frac{\mathcal{N}}{2}\ln\left[\left(\frac{\cos x-\cos\phi}{\cos x-1}\right)^2\right].
\end{equation}
To compute the Hessian determinant as a function of the potential energy per spin, we invert the first case in equation \eqref{eq:e_of_x}, yielding
\begin{equation}
\cos x(v)=1-\frac{1}{2}\exp\left(\frac{2v-1}{\mathcal{N}}\right).
\end{equation}
Inserting this expression into \eqref{eq:f1} gives
\begin{equation}\label{eq:f1withcorrections}
f^{(x(v))}(\phi)=1-2v+\mathcal{N}\ln\left|2(1-\cos\phi)-\exp\left(\frac{2v-1}{\mathcal{N}}\right)\right|.
\end{equation}
As a consequence of the asymptotic behavior \eqref{eq:Nasymptotic} of $\mathcal{N}$, the logarithmic term on the right hand side vanishes and we obtain
\begin{equation}
f^{(x(v))}(\phi)=1-2v.
\end{equation}
From equation \eqref{eq:Szego}, the large-$N$ asymptotic behavior of the Hessian determinant \eqref{eq:Hthetan} is found to be
\begin{equation}\label{eq:1-2e}
\mathcal{D}_1(v):=\lim_{N\to\infty}\bigl|\det\mathcal{H}_N(\theta^{(x)})\bigr|^{1/N}=1-2v
\end{equation}
in the case of $\alpha=1$, valid for accessible potential energy values from the interval $[0,1/2]$. The straight line \eqref{eq:1-2e} is plotted in Fig.\ \ref{fig:D1_of_e} together with numerical results for the Hessian determinants for several finite system sizes. The convergence of the finite-system data to their infinite-system limit $\mathcal{D}_1$ is slow, but this is no surprise as the finite-$N$ corrections in \eqref{eq:f1withcorrections} are logarithmic.

Interpreting \eqref{eq:1-2e} in terms of the criterion of Sec.\ \ref{sec:phasetransitions} [and equation \eqref{eq:criterion} in particular], we observe a vanishing Hessian determinant at the potential energy per spin $v=1/2$. This value coincides precisely with the known phase transition potential energy of the model and nicely illustrates the criterion of Sec.\ \ref{sec:phasetransitions}.

\subsubsection{Exponents $\alpha\in2\NN_0$}
In the case $\alpha\in2\NN_0$ we can use the identity
\begin{equation}
\sum_{j=1}^\infty \frac{\cos(jx)}{j^\alpha}=-(-1)^{\alpha/2}\frac{1}{2}\frac{(2\pi)^\alpha}{\alpha!}B_\alpha\!\left(\frac{x}{2\pi}\right)
\end{equation}
(formula 1.443.1 of \cite{GradshteynRyzhik}), valid for $0\leqslant x\leqslant2\pi$. From this formula, and considering that $x,\phi\in[0,2\pi]$, we can write
\begin{multline}
\sum_{j=1}^\infty \frac{\cos[j(x\pm\phi)]}{j^\alpha}=-(-1)^{\alpha/2}\frac{1}{2}\frac{(2\pi)^\alpha}{\alpha!}\\
\times
\begin{cases}
B_\alpha\!\left(\frac{x\pm\phi}{2\pi}\right) & \text{for $0\leqslant x\pm\phi\leqslant2\pi$},\\
B_\alpha\!\left(\frac{x\pm\phi}{2\pi}\mp1\right) & \text{else}.
\end{cases}
\end{multline}
Inserting these identities into \eqref{eq:fofx}, we obtain
\begin{widetext}
\begin{equation}\label{eq:f2N}
f^{(x)}(\phi)=-(-1)^{\alpha/2}\frac{\mathcal{N}}{2}\frac{(2\pi)^\alpha}{\alpha!}
\begin{cases}
B_\alpha\!\left(\frac{x+\phi}{2\pi}\right)-2B_\alpha\!\left(\frac{x}{2\pi}\right)+B_\alpha\!\left(\frac{x-\phi}{2\pi}\right) & \text{for $\phi\leqslant x$ and $\phi\leqslant 2\pi-x$},\\
B_\alpha\!\left(\frac{x+\phi}{2\pi}-1\right)-2B_\alpha\!\left(\frac{x}{2\pi}\right)+B_\alpha\!\left(\frac{x-\phi}{2\pi}\right) & \text{for $\phi\leqslant x$ and $\phi> 2\pi-x$},\\
B_\alpha\!\left(\frac{x+\phi}{2\pi}-1\right)-2B_\alpha\!\left(\frac{x}{2\pi}\right)+B_\alpha\!\left(\frac{x-\phi}{2\pi}+1\right) & \text{for $\phi> x$ and $\phi>2\pi-x$},\\
B_\alpha\!\left(\frac{x+\phi}{2\pi}\right)-2B_\alpha\!\left(\frac{x}{2\pi}\right)+B_\alpha\!\left(\frac{x-\phi}{2\pi}+1\right) & \text{for $\phi> x$ and $\phi\leqslant 2\pi-x$}.\\
\end{cases}
\end{equation}
For our purposes, the case $\alpha=2$ is particularly interesting, as this is the only positive even exponent for which the $XY$ chain with power law interactions exhibits a phase transition. In this case, making use of the Bernoulli polynomial $B_2(x)=x^2-x+1/6$, the symbol simplifies to
\begin{equation}\label{eq:f2}
f^{(x)}(\phi)=-\frac{6}{\pi^2}
\begin{cases}
\phi^2 & \text{for $\phi\leqslant x$ and $\phi\leqslant 2\pi-x$},\\
(\phi-2\pi)^2-2\pi(x-\phi) & \text{for $\phi\leqslant x$ and $\phi> 2\pi-x$},\\
(\phi-2\pi)^2 & \text{for $\phi> x$ and $\phi>2\pi-x$},\\
\phi^2+2\pi(x-\phi) & \text{for $\phi> x$ and $\phi\leqslant 2\pi-x$}.\\
\end{cases}
\end{equation}
\end{widetext}
To compute the Hessian determinant as a function of the potential energy per spin, we invert the function $v(x)$ in \eqref{eq:e_of_x}, yielding
\begin{equation}
x(v)=\pi\left(1\pm\sqrt{1-4v/3}\right)
\end{equation}
in the case of $\alpha=2$, and insert this expression into \eqref{eq:f2}.

\begin{figure}\center
\psfrag{x}{$x$}
\psfrag{f}{$\phi$}
\psfrag{FFFF}{$f^{(x)}$}
\includegraphics[width=0.72\linewidth]{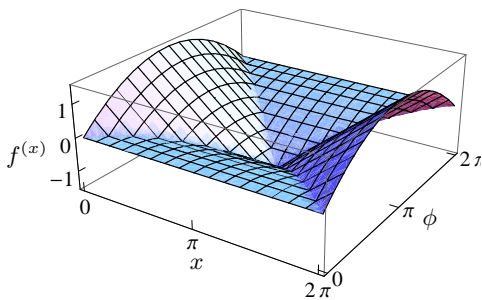}
\caption{\label{fig:f2_of_xphi}
(Color online) The graph of the symbol $f^{(x)}(\phi)$ for $\alpha=2$ as given in \eqref{eq:f2}, plotted as a function of $x$ and $\phi$.
}
\end{figure}

\begin{figure}\center
\psfrag{DDDDDDD}{$\left|\det\mathcal{H}_N\right|^{1/N}$}
\psfrag{e}{$v$}
\includegraphics[width=0.95\linewidth]{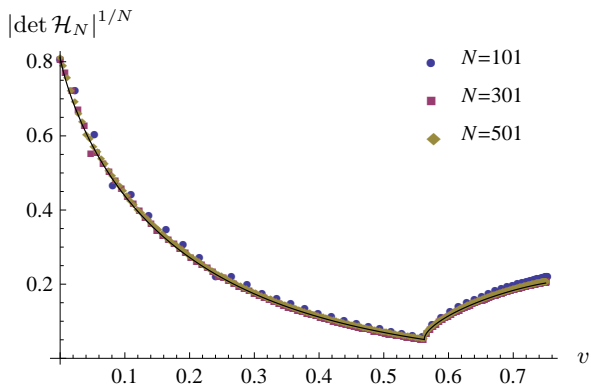}
\caption{\label{fig:D2_of_e}
(Color online) The $N$th root of the Hessian determinant $\det\mathcal{H}_N$ for $\alpha=2$, evaluated at spin wave stationary points $\theta^{(x)}$, plotted versus the corresponding potential energy per spin $v(x)$. Already for moderate system sizes $N$, the finite-system data are in excellent agreement with the analytic large-$N$ asymptotic result $\mathcal{D}_2$ (black line).
}
\end{figure}

The symbol $f^{(x)}(\phi)$, plotted in Fig.\ \ref{fig:f2_of_xphi}, is easily seen to be positive for some values of $x$ and $\phi$, and negative for others. In principle this causes a problem when computing the large-$N$ asymptotic behavior of the Hessian determinant \eqref{eq:Hthetan} from Szeg\"o's theorem \eqref{eq:Szego} where we have to integrate the logarithm of $f^{(x)}$. Instead, to circumvent this problem, we chose to replace $f^{(x)}$ in \eqref{eq:Szego} by its absolute value and compute
\begin{equation}\label{eq:SzegoMod}
\mathcal{D}_2:=\exp\left(\frac{1}{2\pi}\int_0^{2\pi}\ln|f^{(x(v))}(\phi)|\dd\phi\right).
\end{equation}
Inserting \eqref{eq:f2} and performing the integration, we obtain
\begin{widetext}
\begin{equation}\label{eq:D2}
\mathcal{D}_2=\frac{1}{6} \left(3-\sqrt{9-12 v}\right)^2
\begin{cases}
\exp\left[-2+2 \sqrt{2 \sqrt{1-4v/3}-1} \arccoth\left(\sqrt\frac{3-4 v}{2 \sqrt{9-12 v}-3}\right)\right] & \text{for $0\leqslant v\leqslant9/16$},\\[3mm]
\exp\left[-2+2 \sqrt{2 \sqrt{1-4v/3}-1} \arctanh\left(\sqrt\frac{3-4 v}{2 \sqrt{9-12 v}-3}\right)\right] & \text{for $9/16<v\leqslant 3/4$}.
\end{cases}
\end{equation}
\end{widetext}
The graph of $\mathcal{D}_2$ is shown in Fig.\ \ref{fig:D2_of_e} together with numerical results for the Hessian determinants for several finite system sizes. The numerical results are in such excellent agreement with \eqref{eq:D2} that it is tempting to believe that taking the absolute value of $f^{(x(v))}$ in \eqref{eq:SzegoMod} is not merely an approximation, but gives an exact asymptotic expression for the Hessian determinant. Unfortunately, the author was unable to proof this conjecture \footnote{The author suspects a fairly trivial reason for why the replacement of $f^{(x)}$ by its absolute value in \eqref{eq:SzegoMod} works, but is for some reason failing to track it down.}.

Interpreting \eqref{eq:D2} in terms of the criterion \eqref{eq:criterion}, we observe a strictly positive Hessian determinant on the entire range of accessible potential energies per spin $v\in[0,3/4]$. For the class of spinwave stationary points considered, the analysis of the Hessian determinant therefore fails to give an indication of the phase transition known to exist for $\alpha=2$.

\begin{figure}\center
\psfrag{D}{$\mathcal{D}_\alpha$}
\psfrag{e}{$v$}
\includegraphics[width=0.95\linewidth]{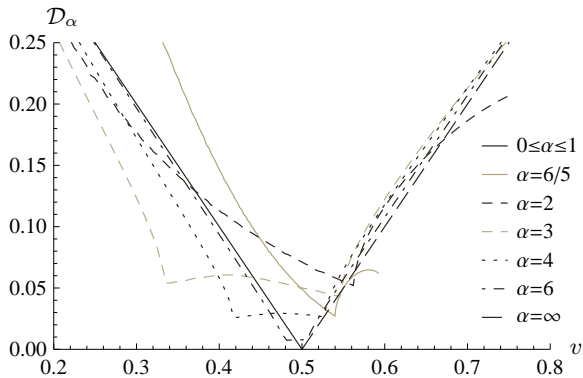}
\caption{\label{fig:Da_of_e}
Asymptotic large-$N$ behavior $\mathcal{D}_\alpha$ of the Hessian determinant, plotted as a function of the potential energy per spin $v$ for various exponents $\alpha$. Dark lines correspond to values of $\alpha$ for which exact analytic results are available, light lines (for $\alpha=6/5$ and $\alpha=3$) were obtained numerically. The dark solid straight line is obtained analytically for $\alpha=1$, but coincides with the numerical results [and also with the Hadamard bound \eqref{eq:bound}] for $\alpha\in[0,1]$. $\mathcal{D}_\alpha$ vanishes at $v=1/2$ for $\alpha\in[0,1]$ (long-range interactions) as well as for $\alpha=\infty$ (nearest-neighbor interactions). For all finite $\alpha>1$, $\mathcal{D}_\alpha$ is bounded away from zero.
}
\end{figure}

For even exponents $\alpha=4$, 6, 8, \dots, analogous calculations can be performed. The corresponding asymptotic results for the determinant, computed according to \eqref{eq:SzegoMod}, share the most important features of the case $\alpha=2$: The analytic large-$N$ asymptotic result $\mathcal{D}_\alpha$ is bounded away from zero (see Fig.\ \ref{fig:Da_of_e}) and in excellent agreement with numerical data (not shown). For $\alpha>2$, however, the fact that $\mathcal{D}_\alpha$ is bounded away from zero was to be expected, as no phase transition occurs in this case. The graphs in Fig.\ \ref{fig:Da_of_e} also suggest that, in the limit $\alpha\to\infty$, $\mathcal{D}_\alpha$ approaches the function $\mathcal{D}_\infty(v)=|2v-1|$. This limit corresponds to nearest-neighbor interactions on the lattice, and indeed $\mathcal{D}_\infty$ coincides with the behavior of the determinant of the nearest-neighbor $XY$ chain reported in Ref.\ \cite{MehtaKastner}. Despite the absence of a phase transition in the $XY$ chain with nearest-neighbor interactions, $\mathcal{D}_\infty$ vanishes at $v=1/2$. Note that this finding is not in conflict with the criterion of Sec.\ \ref{sec:phasetransitions}, since a vanishing $\mathcal{D}$ is not claimed to be sufficient for a phase transition. Moreover, as explained in more detail in Ref.\ \cite{MehtaKastner}, the potential energy $v=1/2$ at which $\mathcal{D}_\infty$ vanishes corresponds to infinite temperatures in the nearest-neighbor model, and at least within the standard canonical setting where temperature is the control parameter, such a transition would be elusive in any case.

\subsubsection{Other values of $\alpha$}
For other values of the exponent $\alpha$, Szeg\"o's theorem can still be used to obtain the asymptotic behavior $\mathcal{D}_\alpha$ of the Hessian determinant but, to the best of the author's knowledge, the infinite sums in \eqref{eq:energy} and \eqref{eq:fofx} cannot be calculated anymore. Numerical results for the cases $\alpha=5/4$ and $\alpha=3$ are shown in Fig.\ \ref{fig:Da_of_e}. Again, $\mathcal{D}_\alpha$ is bounded away from zero for all $\alpha>1$.

\subsection{Hadamard bounds}
The Szeg\"o-type theorem we used in Sec.\ \ref{sec:Szego} allowed us to obtain exact asymptotic large-$N$ results of the Hessian determinant of spinwave stationary points. The drawback, however, is that an evaluation of the resulting infinite Fourier sums is possible only for the exponents $\alpha=1$ and $\alpha\in2\NN_0$. In the present section we will supplement these results by an upper bound on $\mathcal{D}_\alpha$, valid for any $\alpha\geqslant0$. Comparing to numerical data, we will observe that the bound is sharp for $0\leqslant\alpha\leqslant1$.

A bound on the determinant of a (real or complex) $N\times N$-matrix $M$ can be obtained by the celebrated Hadamard inequality
\begin{equation}\label{eq:Hadamard}
|\det M|\leqslant \prod_{j=1}^N \left\lVert c_j\right\rVert,
\end{equation}
where $c_j$ denotes the $j$th column (or row) vector of $M$, and $\lVert c_j\rVert$ its Euclidean norm. In contrast to the methods of Sec.\ \ref{sec:Szego} that are based on the Toeplitz property of the Hessian, the Hadamard inequality can be used to bound the Hessian determinant not only of spinwave stationary points \eqref{eq:isoangular}, but also of any kind of stationary point. In the context of phase transitions and their relation to stationary points and their determinants, Hadamard bounds first have been used in \cite{NardiniCasetti09}.

The Hadamard bound \eqref{eq:Hadamard} becomes particularly simple for a circulant matrix. In this case, $\left\lVert c_j\right\rVert=\left\lVert c_k\right\rVert$ for all $j,k=1,\dots,N$, and hence
\begin{equation}
|\det M|^{1/N}\leqslant \left\lVert c_j\right\rVert
\end{equation}
for any $j$. The Hessian \eqref{eq:Hthetan} of a spinwave stationary point we want to study is not quite a circulant matrix, but it is closely related: $\mathcal{H}_N(\theta^{(x)}$ is an $(N-1)\times(N-1)$-matrix, obtained from an $N\times N$ circulant matrix by deleting one row and one column. The norm of every column of $\mathcal{H}_N(\theta^{(x)}$ is therefore bounded above by the norm
\begin{equation}
\begin{split}
\left\lVert c\right\rVert&=\Biggl(\left[1-2v(x)\right]^2+\mathcal{N}^2\sum_{j=2}^N \frac{\cos^2[x(j-1)]}{\Delta(j-1)^{2\alpha}} \Biggr)^{1/2}\\
&\leqslant\sqrt{[1-2v(x)]^2+\mathcal{N}}
\end{split}
\end{equation}
of any column of the original $N\times N$ circulant matrix. A bound on the Hessian determinant is then given by
\begin{equation}\label{eq:generalbound}
\bigl|\det \mathcal{H}_N(\theta^{(x(v))})\bigr|^{1/N}\leqslant \left\lVert c\right\rVert\leqslant\sqrt{(1-2v)^2+\mathcal{N}}.
\end{equation}
For $\alpha\in[0,1]$, we have observed in \eqref{eq:Nasymptotic} that the normalization constant $\mathcal{N}$ goes to zero in the limit $N\to\infty$, yielding
\begin{equation}\label{eq:bound}
\lim_{N\to\infty}\bigl|\det \mathcal{H}_N(\theta^{(x(v))})\bigr|^{1/N}\leqslant|1-2v|.
\end{equation}
For the accessible values of the potential energy per spin $v\in[0,1/2]$, this bound coincides with the exact $\alpha=1$ asymptotic result for $\mathcal{D}_1$ obtained in Sec.\ \ref{sec:Szego}. Comparing \eqref{eq:bound} to numerical data, it is tempting to conjecture that the bound is tight, i.\,e., coinciding with the exact asymptotics, for all $\alpha\in[0,1]$, but the author was not able to prove this. Such a result for $\mathcal{D}_\alpha$ (i.\,e., one that is independent of the precise value of $\alpha$) would also agree well with the known fact that the thermodynamics of the $XY$ chain has no $\alpha$-dependence as long as $\alpha$ is between zero and one \cite{CampaGiansantiMoroni00}. In particular, the bound \eqref{eq:bound} vanishes at $v=1/2$, and this value coincides with the phase transition potential energy of the model for $\alpha\in[0,1]$. At the same time, the value $v=1/2$ is also the maximum potential energy per spin of the model, and the phase transition is of the ``partial equivalence of ensembles''-type as described in \cite{CaKa07}. For $\alpha>1$, $\mathcal{N}$ converges to the finite value $1/(2\zeta(\alpha))$ in the thermodynamic limit, and the resulting bound \eqref{eq:generalbound} is strictly positive and cannot give any indication of the phase transition of the model. 

\section{Discussion and conclusions}
\label{sec:discussion}

We have analyzed the stationary points of the potential energy function \eqref{eq:Hamiltonian} of the classical $XY$ chain with power law pair interactions, decaying like $r^{-\alpha}$ with the distance $r$ on the lattice. Computing all stationary points of $V$ seems to be way out of reach, but special classes of stationary points can be constructed. For the class of ``spinwave-type'' stationary points where all differences $\theta_i-\theta_{i-1}$ between neighboring angles are equal, we have analytically computed, in the limit of large system size $N$, the asymptotic behavior $\mathcal{D}_\alpha$ of the Hessian determinant of $V$ as a function of the potential energy per spin $v$. The computation is based on the Toeplitz property of the Hessian and makes use of a Szeg\"o-type theorem. The analytic results have been compared to numerical computations of the Hessian determinants for system sizes of up to $N=501$, and the agreement was found to be excellent.

The motivation behind these calculations is based on a recently discovered relation between phase transitions and the properties of stationary points of classical many-body Hamiltonian functions, as reviewed in Sec.\ \ref{sec:phasetransitions}. According to this relation, a phase transition is signaled by the vanishing of the (suitably scaled) Hessian determinant of $V$, evaluated along a suitably chosen sequence of stationary points of $V$, in the thermodynamic limit \eqref{eq:criterion}. Moreover, the thermodynamic limit value of the (potential) energy of such a sequence of stationary points coincides with the phase transition (potential) energy of the model described by $V$.  

For the $XY$ chain with power law pair interactions with exponent $0\leqslant\alpha\leqslant1$, we found that the asymptotic value $\mathcal{D}_\alpha$ of the Hessian determinant at the spinwave stationary points is zero at the potential energy per spin $v=1/2$. In agreement with the criterion on phase transitions and stationary points, this value coincides precisely with the phase transition potential energy of the model.

For $\alpha>1$, $\mathcal{D}_\alpha$ is bounded away from zero, giving no indication of the phase transition occurring for $1<\alpha\leqslant2$. This is of course a somewhat disappointing result, as this is the most interesting case: For $\alpha>1$ an exact solution of the thermodynamics of the $XY$ chain is not known, and obtaining an exact expression for the critical energy would have been a remarkable result. However, the reader should keep in mind that we have considered only the special class of spinwave stationary points. One good reason to focus on this class was the observation that, for the $XY$ chain with nearest-neighbor interactions studied in Ref.\ \cite{MehtaKastner}, the spinwave stationary points were the ``flattest'' ones (in the sense of having the smallest value of $\mathcal{D}_\alpha$ for a given value of the potential energy per spin $v$), and therefore good candidates for $\mathcal{D}_\alpha$ to vanish. On the other hand, the presence of spinwave stationary points depends crucially on the boundary conditions, and their number grows slower than exponentially with the system size. Hence, in order to find an asymptotically vanishing Hessian determinant in the sense of \eqref{eq:criterion}, it appears to be necessary to go beyond the study of spinwave stationary points.

Beyond the analysis of specific features of the $XY$ chain with power law interactions, the results reported in the article provide a number of more general indications that might prove useful for further applications of the criterion of Sec.\ \ref{sec:phasetransitions}: First, the strategies of how to construct special classes of stationary points can be extended straightforwardly to other types of spin-spin-interactions and to higher-dimensional lattices. Second, Szeg\"o-type results should be applicable for the computation of the large-$N$ asymptotics of the Hessian determinant also in other one-dimensional models. And third, the crucial step for successfully applying the criterion to some model is certainly the choice of a suitable class of stationary points. The present case study, although not fully conclusive by itself, provides a further piece of information that can contribute toward an understanding of this issue.

An interesting open question is why the spinwave stationary points discussed in this article allow one to successfully detect the phase transition for exponents $\alpha$ between zero and one, but fail in the case of $\alpha\in(1,2]$, but unfortunately the author can only speculate about the reasons. In general, one would expect that a phase transition can be detected successfully only from stationary points that are somehow ``relevant'' for the transition, in the sense that they correspond to states that dominate the behavior of a phase in the thermodynamic limit. Physical intuition about the phases of a model should therefore be of help when choosing a class of stationary points. But for the $XY$ model with power law interactions, no significant difference seems to distinguish the cases $\alpha\in[0,1]$ and $\alpha\in(1,2]$: In both cases, the transition separates a ferromagnetically ordered from a para\-magnetic phase, and it remains unclear to the author why spinwave stationary points capture the transition in one case, but not the other.

\acknowledgments
The author would like to thank Cesare Nardini for contributing to the computation of the Hadamard bound \eqref{eq:generalbound}. Financial support by the {\em Incentive Funding for Rated Researchers}\/ programme of the National Research Foundation of South Africa is gratefully acknowledged.

\appendix

\bibliography{1dXYalpha.bib}

\end{document}